Original Research Paper

# Multiview Hierarchical Agglomerative Clustering for Identification of Development Gap and Regional Potential Sector

**[1]Tb Ai Munandar, [2]Azhari, [2]Aina Musdholifah and [3]Lincolin Arsyad**

[1]*Department of Informatics Engineering, Faculty of Information Technology, Universitas Serang Raya, Banten, Indonesia*
[2]*Department of Computer Science and Electronic, Math. and Natural Science Faculty*
 *Universitas Gadjah Mada, Yogyakarta, Indonesia*
[3]*Department of Economics, Faculty of Economics and Business, Universitas Gadjah Mada, Yogyakarta, Indonesia*

**Abstract:** The identification of regional development gaps is an effort to see how far the development conducted in every District in a Province. By seeing the gaps occurred, it is expected that the Policymakers are able to determine which region that will be prioritized for future development. Along with the regional gaps, the identification in Gross Regional Domestic Product (GRDP) sector is also an effort to identify the achievement in the development in certain fields seen from the potential GRDP owned by a District. There are two approaches that are often used to identify the regional development gaps and potential sector, Klassen Typology and Location Quotient (LQ), respectively. In fact, the results of the identification using these methods have not been able to show the proximity of the development gaps between a District to another yet in a same cluster. These methods only cluster the regions and GRDP sectors in a firm cluster based on their own parameter values. This research develops a new approach that combines the Klassen, LQ and hierarchical agglomerative clustering (HAC) into a new method named multi view hierarchical agglomerative clustering (MVHAC). The data of GRDP sectors of 23 Districts in West Java province were tested by using Klassen, LQ, HAC and MVHAC and were then compared. The results show that MVHAC is able to accommodate the ability of the three previous methods into a unity, even to clearly visualize the proximity of the development gaps between the regions and GRDP sectors owned. MVHAC clusters 23 districts into 3 main clusters, they are; Cluster 1 (Quadrant 1) consists of 5 Districts as the members, Cluster 2 (Quadrant 2) consists of 12 Districts and Cluster 3 (Quadrant 4) consists of 6 Districts.

**Keywords:** Klassen, LQ, MVHAC, Development Gaps, Potential Sector

## Introduction

Development gap is a global issue faced by many countries in the world, including Indonesia. The development gaps occurred in a region will affect the prioritization of future development. It may occur due to different achievement in development between one region to another. The data from World Bank (2014) states that Indonesia is a country with the most rapidly increasing gaps among the countries in East Asia. Therefore, the policymakers always have measurements on the development they have achieved to see how far the results of the development affect the development gaps between regions.

There are several approaches used to identify the development gaps of a region, such as Klassen typology and Location Quotient (LQ) (Kuncoro and Idris, 2010; Barika, 2012). Klassen Typology divides an area into four quadrants of economic growth pattern (Barika, 2012). Quadrant I (Q1) is the advanced and rapidly growing sector; Quadrant II (Q2) is the advanced but depressed sector; Quadrant III (Q3) is the potential or possible-to-develop sector; and Quadrant IV (Q4) is the relatively underdeveloped sector (Kuncoro and Idris, 2010). However, Klassen Typology is not able to show which sectors are advanced and not that are possessed by a region. Meanwhile, LQ is able to measure and determine the advanced sectors and subsectors of the GRDP sectors owned by a region. Both methods are often used to determine the direction of future regional development policy. However, both methods are used separately in the analysis process that makes the interpretation of development and determination of potential sectors of the region as a whole become difficult.

Some scientists give another approach in identifying the development gaps of the regions, for example, by using the clustering techniques such as Medoid (Spicka, 2013), K-means (Soares *et al.*, 2003; Lukovics, 2009; Bakaric, 2005; Poledníková, 2014) and hierarchical agglomerative clustering (Poledníková, 2014; Kronthaler, 2003; Jaba *et al.*, 2009; Vincze and Mezei, 2011; Vydrová and Novotna, 2012; Nosova, 2013). Nevertheless, the results of the cluster cannot give a strong analytical results yet, that do not have a certain label for each cluster formed. This is what makes the cluster of regional gaps are difficult to interpret, except by those who are already trained.

The basic idea of this research is to develop a new approach to address the inability of Klassen, LQ and clustering techniques in identifying the regional development gaps and determining the potential sector. The concept of multiview clustering used in this research is to combine the ability of these three methods into a single unit. Multiview cluster by agglomerative hierarchical clustering (HAC) is not a new method. Some previous studies have developed the HAC technique into multiview forms as conducted by (Bickel and Scheffer, 2004; Fernandez and Gomez, 2008; Mirzaei, 2010). However, the outcome of the formed clusters still do not have sufficient lable cluster, so that they are difficult to interpret, although visually, the form of multiview clusters already exists.

The combination of these three methods is later called as multiview agglomerative hierarchical clustering (MVHAC). This method is to be able to classify the regions into four major clusters based on the Klassen's rules, to identify the gaps between regions based on the clusters, to identify potential sectors of each region and to visualize the proximity of gaps of between regions against its potential sectors.

### Identification of Regional Development Gaps Using Klassen

The previous research had identified regional development gaps based on the GRDP sectors of the Districts by using several approaches, one of which is by using Klassen typology, as conducted by (Hariyanti and Utha, 2016; Suwandi, 2015; Endaryanto *et al.*, 2015; Fattah and Rahman, 2013; Karsinah *et al.*, 2016). The GRDP sectors are including agriculture, livestock, forestry, fisheries (S1); mining and quarry (S2); manufacture industry (S3); electricity, gas and water supply (S4); construction (S5); trading, hotels and restaurants (S6); transportation and communication (S7); financial, real estate and business services (S8); and other services (S9).

Klassen classification uses the data of GDRP sectors and then classifies the data into four groups of gaps that indicate the level of achievement of development based on the values of Gross Regional Domestic Product (GRDP) sectors. The four groups are; Quadrant I for the advanced and rapidly growing sector; Quadrant II for the advanced but depressed sector; Quadrant III for the potential or possible-to-develop sector; and Quadrant IV for the relatively underdeveloped sector (Kuncoro and Idris, 2010). The Klassen clustering is only able to classify the data of GRDP sectors into four firm groups that have been determined based on the value of the interval value of the growth rate of development and the contribution rate of development between regions compared to the comparison regions. The growth rate of development is calculated based on the equation (1) while the contribution rate of development is calculated based on the equation (2). At Klassen typology, there is a possibility for the values of the growth and contribution rates of the development to have very close intervals, as it is not considered as something that can also provide new information to categorize the data members in one quadrant into new clusters. The illustration of the gaps clusters using Klassen is shown in Fig. 1:

$$r = \frac{P_t - P_{t-1}}{P_{t-1}} \times 100\% \qquad (1)$$

$$y = \frac{P_t + P_{t-1}}{T_t + T_{t-1}} \times 100\% \qquad (2)$$

Where:
$r$ = Growth rate of the development (District and Province)
$y$ = Contribution of the development (District and Province)
$T_t$ = Total GRDP values of all indicators in the observation year
$T_{t-1}$ = Total GRDP values of all indicators in the previous year
$P_t$ = Current GRDP sectors
$P_{t-1}$ = GRDP sectors in the previous year

### Identification of Potential GRDP Sectors of the Regions Using Location Quotient

In addition to the regional development gaps, the potential of supporting sectors of the development of a region can also be identified in order to support the direction of future development. One of the approaches that is often used is Location Quotient (LQ).

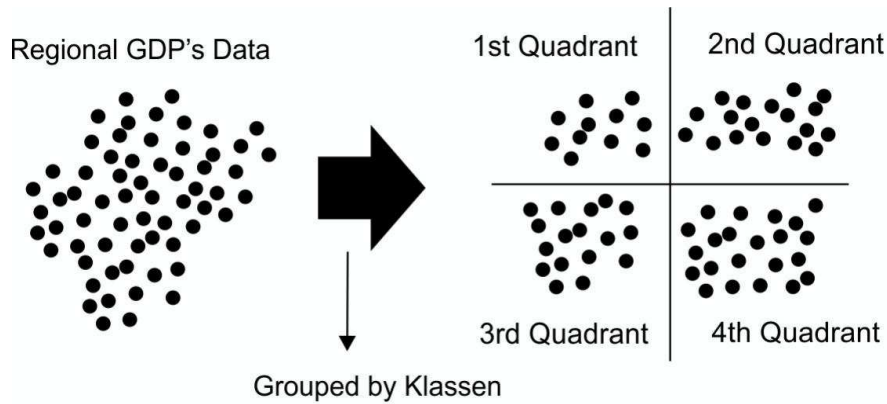

**Fig. 1:** The regional gaps quadrant using Klassen

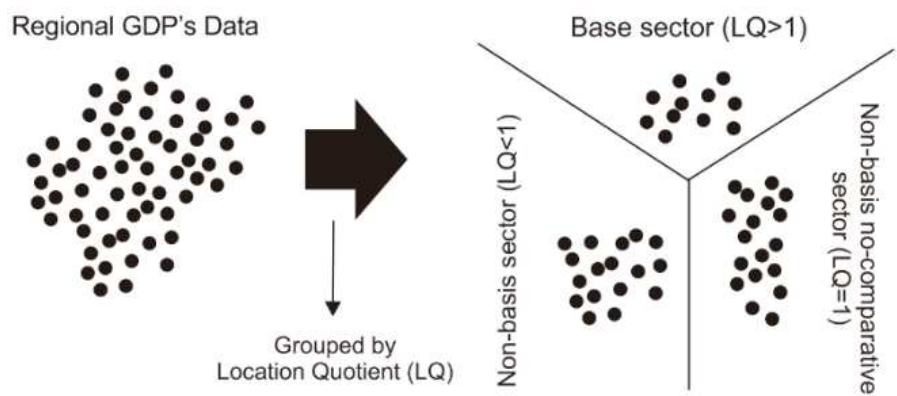

**Fig. 2:** Potential sectors grouping with LQ

LQ is used to illustrate the advantages possessed by a region based on the GDP sectors as conducted by (Kuncoro and Idris, 2010; Bakaric, 2005; Sinaga, 2015). The advanced sectors are typically used to determine which GRDP sectors of a District that would be prioritized for future development and which sectors that should be maintained for the achievement. LQ divides the GDP sectors into three clusters. First, the basic sector with a value of *LQ* >1, means that the sectors become the advanced sectors and are able to meet the needs of the region, even to be exported to meet the needs of other regions. Second, the non-basic sectors with a competitive advantage with a value of *LQ* = 1. These sectors are not only seen to be able to meet the needs of a region itself. Third, the non-basic sectors that do not have a competitive advantage, means that these sectors are considered as the non-potential sectors and still need to be developed, even to be improved continuously in order to be the potential sectors in the future. The calculation of LQ value is conducted by using Equation (3):

$$LQ = \frac{S_i / S}{Y_j / Y} \qquad (3)$$

where, *LQ* is the value of location quotient of each analyzed GRDP sector as well as to show the type of LQ clusters as in the discussion in previous paragraph. $S_i$ is the -*i* added value of GRDP sectors owned by a District; *S* is the -*i* added value of GRDP sectors owned by a Province; $Y_j$ is the -*j* total GRDP of a district and *Y* is total GRDP of a Province. Visually, it is shown in Fig. 2.

*Hierarchical Agglomerative Clustering (HAC) Method*

Agglomerative clustering is data clustering by using bottom-up mode, in which the single-element cluster is clustered based on the closest distance. The graphical result of agglomerative cluster is usually called as dendrogram. Visually, dendrogram shows us on how difficult it combines the two clusters. Clustering is started by calculating the distance between data objects using Equation (4), then each distance of the data is seen as a singleton cluster that will be combined to form a new cluster. The clusters resulted from singleton cluster combination are then clustered again until having the main cluster formed (Sembiring *et al.*, 2010; Saad *et al.*, 2012):

$$D_{euc}(x, y) = \sqrt{\sum_{i=1}^{n} (x_i - y_i)^2} \qquad (4)$$

where, $D_{euc}$ is the eucledian distance between two data objects of $x$ and $y$; $n$ is the number of data dimension; $x$ and $y$, respectively, is the first and second data object that will be calculated for the distance.

*Proposed Method: Multi View Hierarchical Agglomerative Clustering (MVHAC)*

Multiview clustering conducted in this paper refers to the illustration of the concept of multi-cluster conducted by Muller *et al.* (2012). In their paper, the data was clustered into n types of main cluster, then, each member of the main cluster was re-clustered into other cluster types. This basic concept is the concept used in the discussion of this paper related to the development of multiview agglomerative hierarchical clustering (MVHAC).

In this paper, MVHAC is a new approach that is developed for the needs to identify development gaps and regional potential sectors based on the GRDP sectors possessed. MVHAC itself is a combination between HAC, Klassen and LQ clustering techniques as an effort to perfect what cannot be performed by Klassen, LQ and HAC in identifying the regional development gaps. HAC algorithm is chosen as the basis for MVHAC algorithm due to its ability to form the data into clusters, so that it will be possible to see how the two pieces of data combined into a cluster and having a proximity based on the distance. MVHAC algorithm is shown in Fig. 3.

Line 1-6 are the inputs that should be prepared before the clustering process. The inputs used are the data of GRDP sectors of a District in the current analysis year (P) and the previous year (Q). In addition to the GRDP sectors of a District, there are also the GRDP sectors of a Province in the current analysis year (R) and the previous year (S).

| | Algorithm 1: MVHAC |
|---|---|
| 1 | **Input** : |
| 2 | $E = \{P,Q,R,S\}$; (sets of data of PDRB sectors) |
| 3 | $P = \{p_1,p_2,....,p_n\}$; (sets of data of PDRB sectors in current year) |
| 4 | $Q = \{q_1,q_2,....,q_n\}$; (sets of data of PDRB sectors in the previous year) |
| 5 | $R = \{r_1,r_2,....,r_n\}$; (sets of data of PDRB sectors of a Province in current year) |
| 6 | $S = \{s_1,s_2,....,s_n\}$; (sets of data of PDRB sectors of a Province in the previous year) |
| 7 | **Output** : |
| 8 | $DK = \{e_1,e_2,....,e_n\}$; (Growth rate of the development in GRDP sectors of a District) |
| 9 | $DP = \{f_1,f_2,....,f_n\}$; (Contribution of the development in GRDP sectors of a District) |
| 10 | $CK = \{g_1,g_2,....,g_n\}$; (Growth rate of the development in GRDP sectors of a Province) |
| 11 | $CP = \{h_1,h_2,....,h_n\}$; (Contribution of the development in GRDP sectors of a Province) |
| 12 | $K = \{E_{(i)} \mid i = 1,2,....,n\}$; (Development Quadrant Consists of data of GRDP sectors) |
| 13 | *Klabel* = {Quadrant I, Quadrant II, Quadrant III, Quadrant IV}; |
| 14 | $LQ = \{l_1, l_2, l_3, ..., l_n\}$;(LQ value of the data of GRDP sectors) |
| 15 | *Lbl* = {basis, non-basis LQ=1,non-basis LQ<1}; (label of GRDP sectors) |
| 16 | |
| 17 | // calculating the growth and contribution rates of the development (Klassen typology) |
| 18 | **Foreach** *input* $p_i \in P$, $q_i \in Q$, $r_i \in R$, $s_i \in S$; $1 \leq i \leq n$ **do** |
| 19 | $DK(p_i,q_i) \leftarrow sqrt(sqr(sum(p_i - q_i)))$; |
| 20 | $DP(r_i,s_i) \leftarrow sqrt(sqr(sum(r_i - s_i)))$; |
| 21 | $CK(p_i,q_i) \leftarrow sqrt(sqr(sum(p_i - q_i)))$; |
| 22 | $CP(r_i,s_i) \leftarrow sqrt(sqr(sum(r_i - s_i)))$; |
| 23 | |
| 24 | //The form of main cluster 4 with Klassen's rules |
| 25 | **Foreach** $e_i \in DK$, $f_i \in DP$, $g_i \in CK$, $h_i \in CP$; $1 \leq i \leq n$ **do** |
| 26 | **If** $(e_i \geq f_i)$ *and* $(g_i \geq h_i)$ **then** |
| 27 | $K \leftarrow \{Ei\}$; |
| 28 | *Klabel* $\leftarrow$ Quadrant I; |
| 29 | **Elseif** $(e_i < f_i)$ *and* $(g_i \geq h_i)$ **then** |
| 30 | $K \leftarrow \{Ei\}$; |
| 31 | *Klabel* $\leftarrow$ Quadrant II; |
| 32 | **Elseif** $(e_i \geq f_i)$ *and* $(g_i < h_i)$ **then** |
| 33 | $K \leftarrow \{Ei\}$; |
| 34 | *Klabel* $\leftarrow$ Quadrant III; |
| 35 | **Else** |
| 36 | $K \leftarrow \{Ei\}$; |
| 37 | *Klabel* $\leftarrow$ Quadrant IV; |
| 38 | **Endif** |
| 39 | **End** |
| 40 | **Foreach** *input* $p_i \in P$ *and* $r_i \in R$ $\{1 \leq i \leq n\}$ **do** |
| 41 | $LQ \leftarrow (p_i / r_i)/(sum(p_i)/sum(r_i))$; (calculation of Location Quotient value) |
| 42 | **Foreach** $l_i \in LQ$ **do** |
| 43 | **If** $l_i > 1.0$ **then** |
| 44 | $K \leftarrow LQ_i$; |
| 45 | *lbl* $\leftarrow$ basis; |

```
46                          Elseif lᵢ = 1.0 then
47                              K ← LQᵢ;
48                              lbl ← non-basis LQ=1;
49                          Else
50                              K ← LQᵢ;
51                              lbl ← non-basis LQ<1;
52                          Endif
53                      End
54                      Calculate the matrix of distance D for every object that has an LQ value for all pᵢ ∈ P in K;
55                      Define the set of clusters based on singleton cluster, in which every set of clusters is the representation of every pᵢ ∈
56                      P in K;
57
58                      // The integration of singleton cluster in each main cluster
59                      t ← 0;
60                      Repeat
61                              t ← t+1;
62                              // Integrate 2 single clusters pᵢ ∈ P from the set of clusters K, using the following ways :
63                              // (if D uses Single Linkage)
64                              D(p₁,p₂) ← min(Distance(p₁,p₂));
65                              // (if D uses Complete Linkage)
67                              D(p₁,p₂) ← max(Distance(p₁,p₂));
68                              // (if D uses Average Linkage)
69                              D(p₁,p₂) ← (1/(nₐ.nᵦ)).sum(sum(Distance((p₁,p₂)));
70                              // (if D uses Centorid Linkage)
71                              D(p₁,p₂) ← ||p̄₁ − p̄₂ ||₂;
73                              // Update the matrix of proximity with the new distance between the recently formed cluster and the
74                              origin cluster.
75                              delete p₁ and p₂ from K
76                              add { p₁ , p₂ } into K
77                      Until cluster K = 1;
78                  End
79          End
80  End
81
```

**Fig. 3:** MVHAC Algorithm

Line 7-15 are the outputs resulted from the inputs in Line 1-6. The outputs are the growth rate of the development in GRDP sectors of a District (DK), the contribution of the development in GRDP sectors of a District (DP), the growth rate of the development in GRDP sectors of a Province (CK), the contribution of the development in GRDP sectors of a Province, Development Quadrant consists of the data of GRDP sectors (K) and the LQ value of the data of GRDP sectors (LQ).

The calculation of the growth rate of the development of a District was conducted in Line 19, while for a Province was conducted in Line 20. The growth rate was obtained based on the Equation (1). Line 21 is the calculation of the contribution of the development of a District, while for a Province was calculated in Line 22. The calculations in Line 21-22 were conducted by using Equation (2).

In this algorithm, the first clustering was conducted on Line 25-39 by using Klassen's rules. This clustering resulted in four main clusters in which every District would be clustered into Quadrant I, II, III and IV based on the comparison of growth rate and contribution of the development that have been calculated in Line 19-22. Line 40-53 consist of clustering process of GRDP sectors owned by every District that have been clustered previously based on the process in Line 25-39. This second-stage of clustering process was started by calculating the Location Quotient (LQ) value of each GRDP sector owned by each District towards all Quadrants.

The Multiview clustering process was conducted in Line 54-81. In this stage, every GRDP sector possessed by each District will be clustered towards all Districts in each Quadrant. The clustering in this stage was started by calculating the distance between on GRDP sector to another in each Quadrant. The calculation of distance between GRDP sectors was conducted by using Equation (3). After calculating all distance data, the clustering was conducted on 2 data of GRDP sectors with the closest distance.

The visualization of Multiview cluster in this discussion is shown in Fig. 4. Figure 4 shows the results of first clustering using Klassen's rules, so that Cluster n (Quadrant n) has regional 1, regional 2, until regional n as its members. Every regional has GRDP sectors which are then calculated for the Location Quotient (LQ) value for re-clustering based on the closest distance possessed by the GRDP sectors of a District in related cluster.

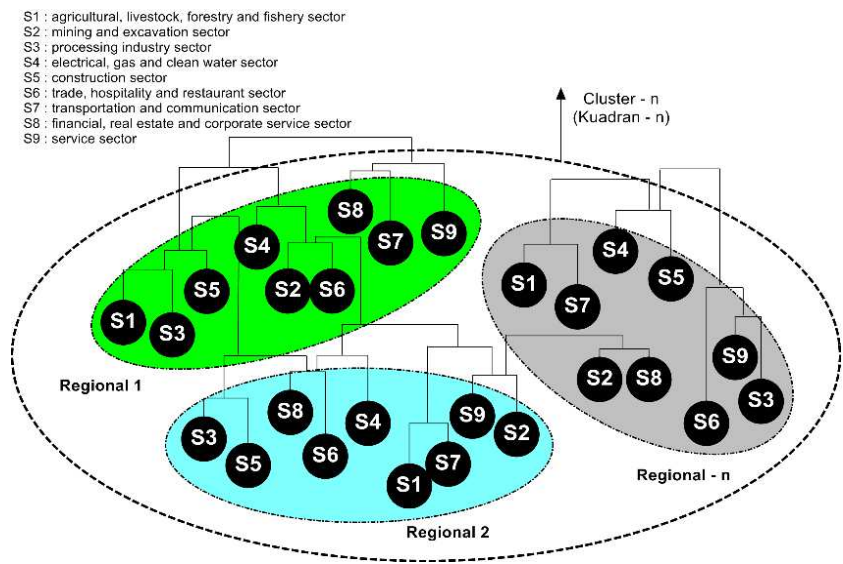

**Fig. 4:** Visualization of clustering using MVHAC

**Table 1:** The clusters of regional gaps based on Klassen

| No. | District | Quadrant | No. | District | Quadrant |
|---|---|---|---|---|---|
| 1 | Kab. Garut | K2 | 13 | Kab. Cirebon | K2 |
| 2 | Kab. Indramayu | K2 | 14 | Kab. Karawang | K2 |
| 3 | Kota Bogor | K4 | 15 | Kab. Sukabumi | K2 |
| 4 | Kab. Majalengka | K4 | 16 | Kota Banjar | K4 |
| 5 | Kab. Purwakarta | K2 | 17 | Kab. Bekasi | K1 |
| 6 | Kab. Tasikmalaya | K4 | 18 | Kab. Cianjur | K2 |
| 7 | Kota Tasikmalaya | K4 | 19 | Kab. Kuningan | K2 |
| 8 | Kota Bekasi | K1 | 20 | Kab. Sumedang | K1 |
| 9 | Kota Bandung | K2 | 21 | Kota Cirebon | K2 |
| 10 | Kab. Bandung | K2 | 22 | Kota Depok | K1 |
| 11 | Kab. Bogor | K2 | 23 | Kab. Subang | K2 |
| 12 | Kab. Ciamis | K4 | | | |

**Table 2:** The Clusters of Gaps in GRDP sectors of the Districts

| District | S1 | S2 | S3 | S4 | S5 | S6 | S7 | S8 | S9 |
|---|---|---|---|---|---|---|---|---|---|
| Kab. Garut | 1 | -1 | -1 | -1 | -1 | 1 | -1 | -1 | -1 |
| Kab. Indramayu | 1 | 1 | -1 | -1 | -1 | 1 | -1 | -1 | -1 |
| Kota Bogor | -1 | -1 | 1 | -1 | -1 | 1 | -1 | -1 | -1 |
| Kab. Majalengka | 1 | -1 | -1 | -1 | -1 | 1 | -1 | -1 | -1 |
| Kab. Purwakarta | -1 | -1 | 1 | -1 | -1 | 1 | -1 | -1 | -1 |
| Kab. Tasikmalaya | 1 | -1 | -1 | -1 | -1 | 1 | -1 | -1 | -1 |
| Kota Tasikmalaya | -1 | -1 | 1 | -1 | -1 | 1 | -1 | -1 | -1 |
| Kota Bekasi | -1 | -1 | 1 | -1 | -1 | 1 | -1 | -1 | -1 |
| Kota Bandung | 1 | -1 | -1 | -1 | -1 | -1 | -1 | -1 | -1 |
| Kab. Bandung | -1 | -1 | 1 | -1 | -1 | -1 | -1 | -1 | -1 |
| Kab. Bogor | -1 | -1 | 1 | -1 | -1 | 1 | -1 | -1 | -1 |
| Kab. Ciamis | 1 | -1 | -1 | -1 | -1 | 1 | -1 | -1 | -1 |
| Kab. Cirebon | 1 | -1 | 1 | -1 | -1 | 1 | -1 | -1 | -1 |
| Kab. Karawang | -1 | -1 | 1 | -1 | -1 | -1 | -1 | -1 | -1 |
| Kab. Sukabumi | 1 | -1 | -1 | -1 | -1 | 1 | -1 | -1 | -1 |
| Kota Banjar | -1 | -1 | -1 | -1 | -1 | 1 | -1 | -1 | 1 |
| Kab. Bekasi | -1 | -1 | 1 | -1 | -1 | -1 | -1 | -1 | -1 |
| Kab. Cianjur | 1 | -1 | -1 | -1 | -1 | 1 | -1 | -1 | -1 |
| Kab. Kuningan | 1 | -1 | -1 | -1 | -1 | 1 | 1 | -1 | -1 |
| Kab. Sumedang | 1 | -1 | 1 | -1 | -1 | 1 | -1 | -1 | -1 |
| Kota Cirebon | -1 | -1 | -1 | -1 | -1 | 1 | -1 | -1 | -1 |
| Kota Depok | -1 | -1 | 1 | -1 | 1 | 1 | -1 | -1 | -1 |
| Kab. Subang | 1 | -1 | -1 | -1 | -1 | 1 | -1 | -1 | -1 |

*The Comparison of the Identification of Regional Development Gaps Using Klassen, LQ, HAC and MVHAC*

This section discusses the identification process of the development gaps using three different methods, they are; Klassen, LQ, HAC and MVHAC. It is conducted to see and compare the results of those methods, so that it will obtain the better approach. The data of GRDP sectors of 23 Districts in West Java Province of 2011 and 2012 were used to test these three methods.

The first test was conducted by using Klassen approach. In order to test the data in Klassen method, it requires the data of the sectors in current year that will be tested and the data in the previous year that is used as the comparative data. In this first test, the test data is the data in 2012, while the comparative data used is the data in 2011. The results of test of the data of GRDP sectors by using Klassen show that 23 Districts in West Java province are divided into three clusters: Quadrant I (K1), Quadrant II (K2) and Quadrant IV (K4). The clustering of the development gaps using Klassen is shown in Table 1.

Table 1 indicates that there are 5 Districts included into a cluster with advanced and rapidly growing development (K1). 12 Districts are included into a cluster with advanced but depressed development (K2). While the remaining 6 Districts are included into a cluster with a relatively underdeveloped level (K4). The results of this clustering are obtained by comparing the growth rate of the development of the districts and the province and the contribution rate of the development of the province based on its GRDP sectors data. In the clustering using Klassen, every region is clustered firmly into quadrants of development achievement suited to the rules that have been set.

The next test is by using LQ towards the data of GRDP sectors in 2012. The LQ method categorizes the gaps in GRDP sectors based on three clusters based on the LQ value possessed. The results show that most GRDP sectors possessed by the Districts in West Java Province are included into the non-basic clusters that do not have a competitive advantage. Table 2 shows the results of the clustering of GDP sectors of each District using LQ.

Table 2 shows that there are 161 non-basic sectors that do not have a competitive advantage and 46 basic sectors which are the potential possessed by the related districts. The potential sectors are dominated by agriculture, livestock, forestry, fisheries (S1), manufacture industry (S3) and trading, hotels and restaurants (S6). Figure 5 shows the graphich of the distribution gaps of GRDP sectors throughout the Districts in West Java Province.

The test of the GRDP sectors data was then conducted by using the HAC method. In the testing using HAC, the data of GRDP sectors of 23 Districts of West Java Province in 2011 and 2012 were then clustered. The results of the second clustering of the data are shown in Fig. 6 and 7. The output of the cluster is able to show the position of the proximity of one region to another. It can be said that it is a proximity of development achievement since the clustering is based on the GRDP sectors data of each District.

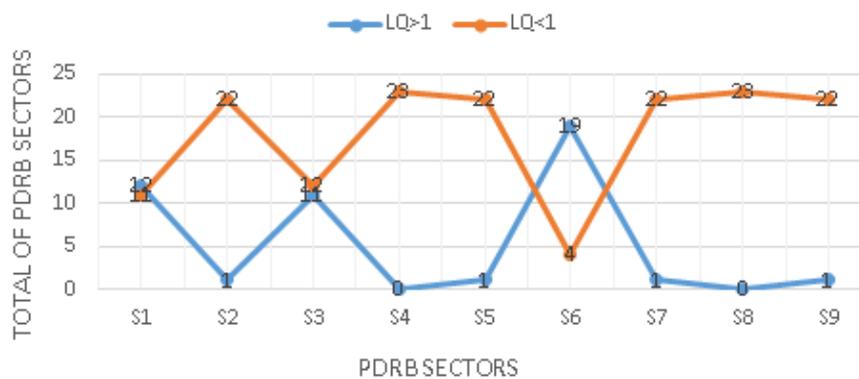

**Fig. 5:** Graphic of the distribution of GRDP sectors clusters using LQ

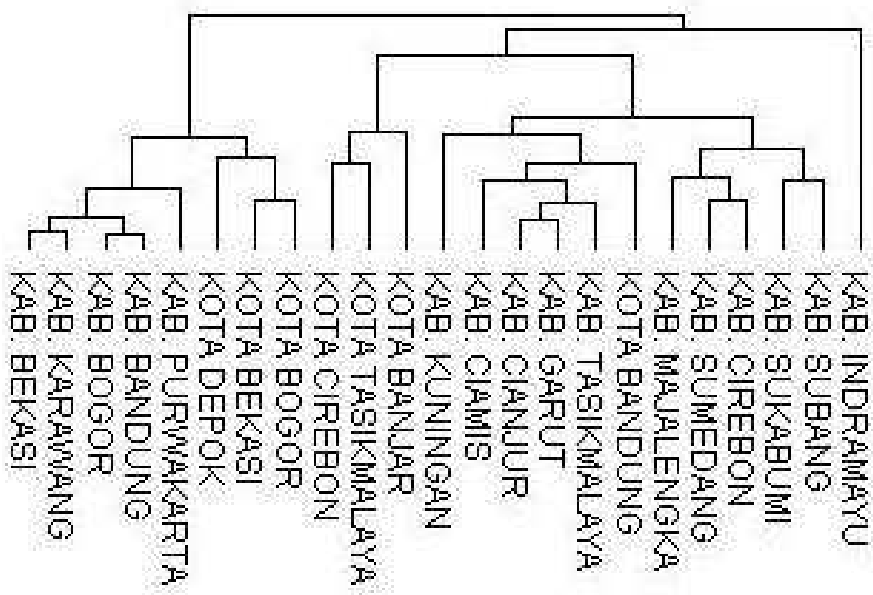

**Fig. 6:** The Results of the clustering of the GRDP sectors data of 2011

The results of the cluster using HAC in Fig. 6 and 7 show that some Districts have a fixed development achievement. Such as Bekasi and Karawangan Districts, Bandung and Bogor District, Bogor City, Bekasi and Depok City, as well as other cities which are still in a close position for development achievements. However, there are some districts that appear to move for their positions based on the results of the data clusters in 2011 and 2012.

The results of the data clusters in 2011, Majalengka has a close development achievement to Cirebon and Sumedang Districts. Meanwhile, the results of the data cluster in 2012, Majalengka has a close development achievement to Subang district. In addition, in a data cluster in 2012, Sukabumi has a close achievement to Subang district, whereas the results of data clusters in 2012 show that Subang District has changed for its position.

In a test using HAC method, the data of GRDP sectors of the Districts are not only clustered hierarchically, but also able to visualize the proximity of the development achievements of one region to another based on the clusters formed. This ability is what cannot be obtained in earlier test using Klassen. Nevertheless, the results of the clusters by HAC are also not able to show the level of development achievement as Klassen does.

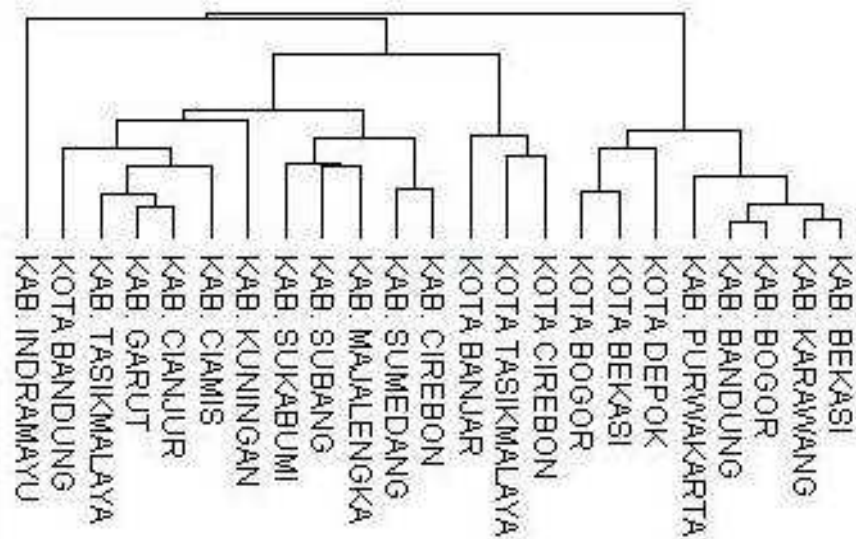

**Fig. 7:** The Results of the clustering of the GRDP sectors data of 2012

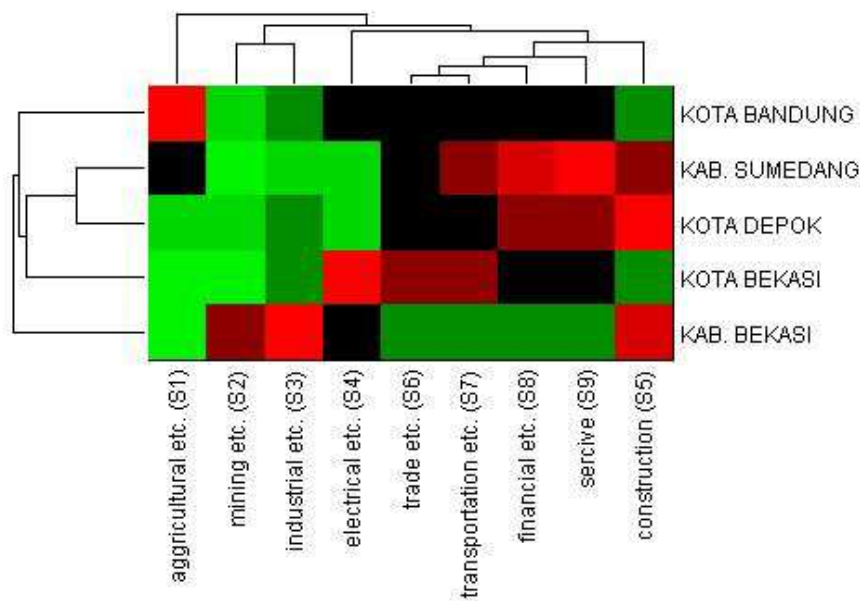

**Fig. 8:** Multiview cluster for Quadrant 1 (cluster 1)

The final test of the data of GRDP sectors was conducted by using the Multi-View Hierarchical Agglomerative Clustering (MVHAC) approach. The data of GRDP sectors in 2011 and 2012 are used to be analyzed by using MVHAC. The test results using MVHAC are shown in Fig. 7-9. MVHAC divides the GRDP sectors of 23 Districts of West Java Province into three main clusters, namely Quadrant 1 (K1), Quadrant 2 (K2) and Quadrant 4 (K4). Each cluster formed shows the cluster of development based on Klassen.

Figure 8 shows the first cluster (K1) with 5 member regions. Each region was then clustered based on Klassen and LQ value of each GRDP. In K1, it is shown that Sumedang District and Depok City form a cluster level 1, which means that the close development gaps. At the next level, cluster level 1 forms the cluster level 2 on Bekasi City. Level 2 indicates that the development gaps of Bekasi City have a close gap with Sumedang District and Depok City. It is similar to other districts as shown in Fig. 8. The results of clustering also show how each district is clustered not only based on the regional data, but also on the value of its GRDP sectors.

The second main cluster, Quadrant 2 (K2), consists of 12 members as shown in Fig. 9. As in the first main cluster, each district is clustered hierarchically not only based on its regional GRDP data, but also on the LQ value of every GRDP sector possessed. It is also applied to six other Districts that are included into the third main cluster (K4) as shown in Fig. 10.

Based on the test on all four approaches to identify the regional development gaps, it is known that the MVHAC method is able to cover some of the functions that are not able to conduct by Klassen, LQ and HAC. In addition to identify the development gaps through the results of the clusters, MVHAC is also able to cluster the Districts based on the LQ value based on the GRDP sectors possessed. Moreover, it is also added by the ability to visualize the proximity of development gaps and potential sectors possessed by every District. Table 3 shows the differences in the ability of the four methods.

**Table 3:** Differences between Klassen, LQ, HAC and MVHAC

| Ability | Klassen | LQ | HAC | MVHAC |
|---|---|---|---|---|
| Regional Development Gaps Clustering | Yes | Yes | No | Yes |
| Graphic Visualization of the Gaps | No | No | Yes | Yes |
| Regional GRDP sectors Clustering | Yes | No | No | Yes |
| Visualization of the proximity of regional development gaps | No | No | Yes | Yes |
| Visualization of the proximity of the achievement of the development in GRDP sectors | No | No | Yes | Yes |

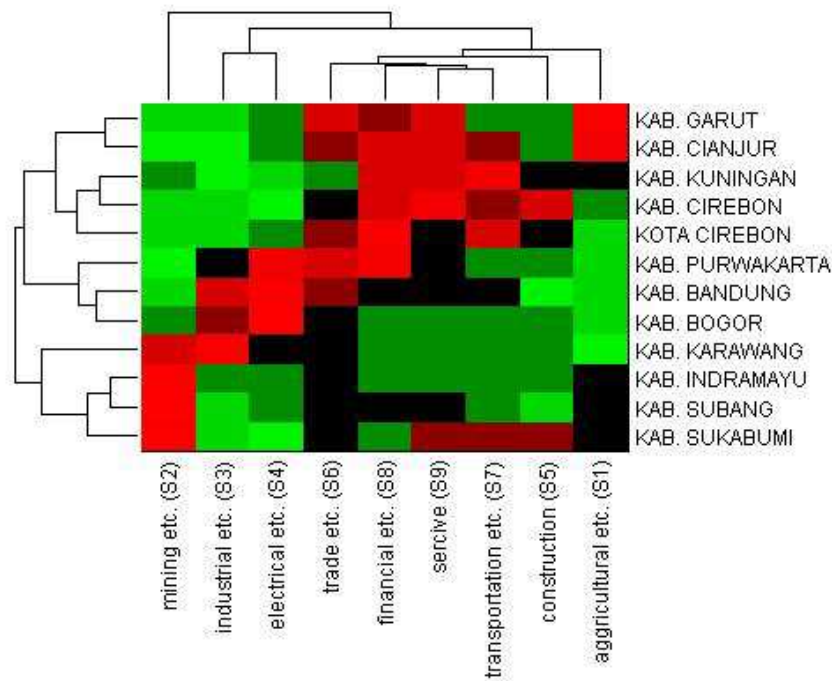

**Fig. 9:** Multiview cluster for Quadrant 1 (cluster 2)

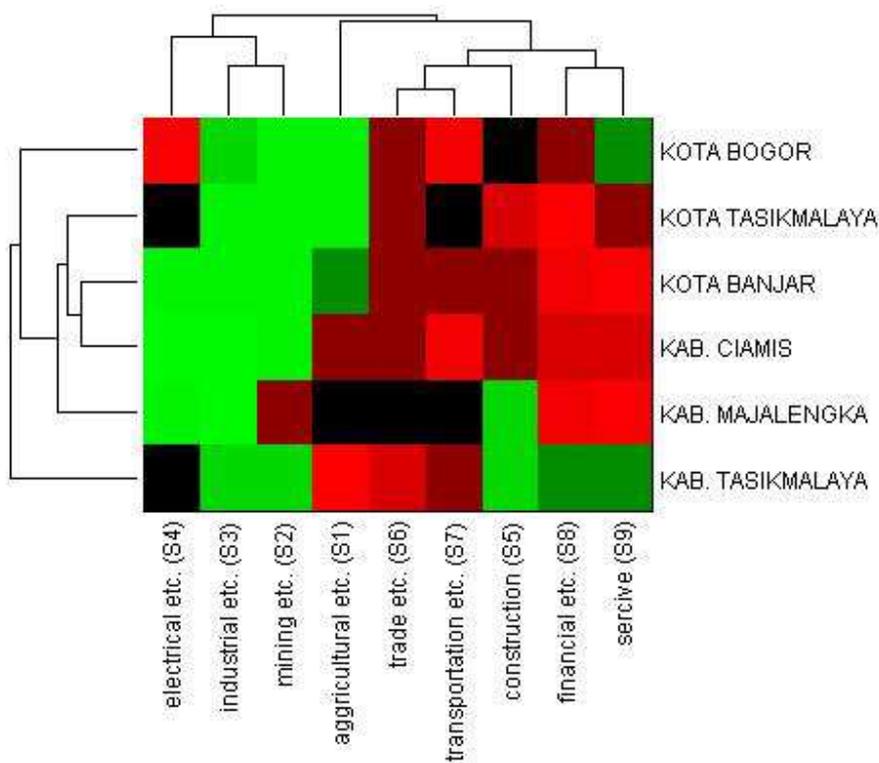

**Fig. 10:** Multiview cluster for Quadrant 4 (cluster 3)

## Conclusion

Based on the research conducted, it is obtained that the MVHAC method can be used to identify the regional development gaps based on the GRDP sectors possessed. Unlike the Klassen, LQ and HAC methods, the MVHAC method is able to combine the abilities of those three methods. In addition, MVHAC is also able to show the proximity of the development gaps occurred in the form of hierarchical clusters. The clusters resulted from MVHAC cluster the districts into three main clusters, they are; Quadrant 1 (Q1), Quadrant 2 (Q2), Quadrant 4 (Q4), in which every member of the main clusters are re-clustered based on the GRDP sectors possessed. There are 5 Districts which show the level of regional development as the advanced and rapidly growing regional development (Cluster 1- K1), 12 Districts show the advanced but depressed development, (Cluster 2-K2) and 6 Districts are the regions with relatively underdevelopment level (Cluster 3-K4).

Although the results of the research are able to classify the PDRB sector data into multiview clusters, the newly formed cluster multiview is limited to the objects in each major cluster. Further development can be made possible that the form of a multiview cluster for the objects between major clusters, so that the analysis of regional development inequality can be more detailed related to differences in development achievements in each major cluster.

## Acknowledgement

Acknowledgments are conveyed to Universitas Serang Raya through Institute for Research and Community Service which has funded this research.

## Author's Contributions

**Tb Ai Munandar:** In this study his most important contribution is data analysis using HAC, LQ and Klassen and test the accuracy of grouping results using MVHAC algorithm.

**Azhari:** In this study his most important contribution is test the accuracy of grouping results using MVHAC algorithm and Review the paper before submitting.

**Aina Musdholifah:** In this study her most important contribution is test the accuracy of grouping results using MVHAC and HAC algorithm and Review the paper before submitting.

**Lincolin Arsyad:** In this study his most important contribution is data analysis using LQ and Klassen and Review the paper before submitting.

## Ethics

No ethical issues would arise after the publication of this manuscript.